\documentclass[twocolumn,showpacs,aps,epsfig,nofootinbib]{revtex4}

\usepackage{graphicx}
\usepackage{epstopdf}
\usepackage{latexsym}

\usepackage[center]{subfigure}

\begin{document}

 \newcommand{\bq}{\begin{equation}}
 \newcommand{\eq}{\end{equation}}
 \newcommand{\bqn}{\begin{eqnarray}}
 \newcommand{\eqn}{\end{eqnarray}}
 \newcommand{\nb}{\nonumber}
 \newcommand{\lb}{\label}
\newcommand{\PRL}{Phys. Rev. Lett.}
\newcommand{\PL}{Phys. Lett.}
\newcommand{\PR}{Phys. Rev.}
\newcommand{\CQG}{Class. Quantum Grav.}

\title{Stability   of spin-0 graviton and strong coupling  in Horava-Lifshitz  theory of  gravity} 

\author{Anzhong Wang}
\email{anzhong_wang@baylor.edu}

\affiliation{ GCAP-CASPER, Physics Department, Baylor
University, Waco, TX 76798-7316, USA }

\author{Qiang Wu}
\email{wuq@zjut.edu.cn}

\affiliation{  Department of Physics, Zhejiang University of
Technology, Hangzhou 310032,  China  }

\date{\today}

\begin{abstract}

In this paper, we consider  two different issues,  stability  and strong coupling,  raised lately in the 
newly-proposed Horava-Lifshitz (HL) theory of  quantum gravity with projectability condition. We 
find that all the scalar modes are stable in the de Sitter background, due to two different kinds 
of effects, one from high-order derivatives of the spacetime curvature, and the other from the 
exponential expansion of the de Sitter space. Combining these  effects  properly, one can make 
the instability found in the Minkowski background never appear even for small-scale modes, provided 
that the IR limit  is  sufficiently closed to  the  relativistic  fixed point. At the fixed point, all the modes 
become stabilized. We also show that the instability of Minkowski spacetime can be  cured by 
introducing  mass to the spin-0 graviton. The strong coupling problem is investigated following the
effective field theory approach, and found that it cannot be cured by the Blas-Pujolas-Sibiryakov 
mechanism, initially designed for the case without projectability condition, but might be circumvented  
by the Vainshtein mechanism, due to the non-linear effects. In fact, we construct a class of exact 
solutions, and show explicitly that it reduces smoothly to the de Sitter spacetime in the relativistic 
limit.

\end{abstract}


\maketitle

\section{Introduction}
\renewcommand{\theequation}{1.\arabic{equation}} \setcounter{equation}{0}

Properly formulating the theory of quantum gravity has been one of the main driving forces
in gravitational physics over past several decades \cite{QGs}. Although there are several
very promising candidates, including Loop Quantum Gravity \cite{LQG} and string/M theory
\cite{strings},   it is fair to see that our understanding of it is still
very limited. Horava recently proposed another alternative \cite{Horava}, motivated by 
the Lifshitz theory in solid state physics \cite{Lifshitz}, for which it is often referred 
to as the Horava-Lifshitz (HL) theory. It has various remarkable features, including
its power-counting renormalizability \cite{OR}, the divergence of its effective speed of light
in the ultra-violet (UV), which could potentially resolve the horizon problem without 
invoking inflation  \cite{KK}. Scale-invariant super-horizon curvature 
perturbations can also be produced without inflation 
\cite{Muka,Piao,GWBR,WM,YKN,WWM}, and dark matter and dark energy 
can have their geometric origins \cite{Mukb,fR}. 
Furthermore, bouncing universe can be easily constructed due to the
high-order derivative terms of the spacetime curvature \cite{calcagni,brand,WW}.
For detail, we refer readers to  \cite{HWW} and references therein.
 
Despite all of these attractive features, the theory plagues with two serious problems:
the instability of the Minkowski background  \cite{Horava,BS,SVW,WM,Kob}, and the strong 
coupling \cite{CNPS,BPSa,KA,PS,KP}. To solve these problems, various modifications
were proposed \cite{fRR,Horava2}. In particular,  Blas, Pujolas and Sibiryakov (BPS) 
\cite{BPSb} found that inclusion of  terms made of $a_{i}$, where
 \bq
 \lb{1.a}
 a_{i} = \partial_{i}\ln(N),
 \eq
can cure the instability of the Minkowski background, where $N$ is  the lapse 
function. Of course, this is possible
only for the version of the HL theory without projectability condition. Otherwise,  $N$ depends only on time, and $a_i$ vanishes
identically.  By properly choosing the coupling constants, the strong coupling problem
can be also addressed in such a setup \cite{BPSc}. The main idea is to introduce two
energy scales, the UV cutoff scale $M_{*}$ and the strong coupling scale $\Lambda_{k}$.
If $M_{*}$ is low enough so that
\bq
\lb{1.b}
M_{*} \lesssim \Lambda_{k},
\eq
then the linear perturbations become invalid before   $\Lambda_{k}$ is
researched, so that  the strong coupling problem does not show up at all [cf. Fig.\ref{fig2}].
Applications of the BPS model  to cosmology were studied recently in \cite{KUY,CB,CHZ}, while 
spherically symmetric spacetimes were investigated in \cite{Kirs}. 
However, a  price to pay in such a setup   is the enormous number of independent 
coupling constants. It can be shown that only the sixth-order derivative terms 
in the potential are more than 60  \cite{KP}. 
It should be also noted that
giving  up the projectability condition often causes the theory to suffer the inconsistence 
problem \cite{LP}.   However, Kluson  recently showed that the Hamiltonian formalism of the BPS model 
is very rich, and that the algebra of constraints is well-defined 
 \cite{Kluson}.

On the other hand, Sotiriou, Visser and Weinfurtner (SVW) generalized the original
version of the HL theory to the  most general form by giving up the detailed balance
condition but still keeping the  projectability one \cite{SVW}. In the SVW setup, the inconsistence
problem does not exist, and the 
gravitational sector contains totally ten coupling constants, $G, \; \Lambda, \; \xi, \;
g_{i}\; (i = 2, 3, ..., 8)$, where $G$ and $\Lambda$ denote, respectively, the 4-dimensional
Newtonian and cosmological constants, $\xi$ and $g_{n}$'s are other coupling
constants, due to the breaking of the Lorentz invariance of the theory. Although 
the Minkowski background is still not stable in such a setup \cite{SVW,WM}, de
Sitter spacetime is  \cite{HWW}. Therefore, in such a setup, one may consider
the latter as its legitimate  background, similar to what happened in the massive
gravity \cite{MGs}. Moreover, the SVW setup also faces the strong coupling problem 
\cite{KA}. Recently, Mukohyama  showed that this problem could be solved 
by the Vainshtein mechanism \cite{Vain}, at least as far as the spherically symmetric, 
static, vacuum spacetimes are concerned \cite{Mukc}.

In this paper, our purposes are two-folds. We first generalize our studies of
\cite{HWW} to include high-order derivative terms, whereby we show explicitly
that all the scalar modes, including the short-scale ones, are stable in
the de Sitter spacetime, by properly combining  two different kinds of effects,  
one from high-order derivatives of the spacetime curvature, and the other from the 
exponential expansion of the de Sitter space. This is  done in Sec. II.
Second, we systematically study the strong coupling problem 
by following the effective theory approach \cite{Pol}, and show clearly that the 
BPS mechanism for solving the strong coupling
problem originally invented in the case without projectability condition cannot be applied to the SVW
case with projectability condition. This is consistent with the results found by
BPS using the St\"uckelberg trick \cite{AGS,RT}. This is done in Sec. III. 
In Sec. IV, we construct a class of non-perturbative cosmological solutions, and show that it 
reduces smoothly  to the de Sitter  spacetime (with rotation)  in the relativistic limit.  This
implies that the spin-0 graviton  indeed decouples in the IR limit and does not cause additional problem, once nonlinear 
effects are included,  a similar situation also happens other theories, such as the DGP model \cite{DDGV} 
\footnote{It should be noted that, although in both theories it is the non-linear interaction that 
makes the theories consistent with observations,  there is a fundamental difference between them:
the DGP model  represents the modification of general relativity in the IR, 
while the HL theory modifies general relativity mainly  in the UV.  So, the  coupling of the scalar graviton 
in the DGP model is of order one, and the non-linear effects help to screen its coupling   to external sources.
In the HL theory, on the other hand, the scalar graviton is self-interacting, and it is this interaction that leads to
the strong coupling problem. See the analysis carried in Sec. III. }.
This can be considered as the generalization of Mukohyama's analysis of the 
spherical case to the cosmological one. In Sec. V, we present our main conclusions, and
shown that  the Minkowski spacetime can also become stable by introducing  mass 
to the spin-0 graviton.

\section{Stability of de Sitter Spacetime}

\renewcommand{\theequation}{2.\arabic{equation}} \setcounter{equation}{0}

To start with, in this section we shall first give a brief introduction to the SVW setup, and then
consider linear perturbations in the de Sitter background.

\subsection{The SVW Setup}

 Sotiriou, Visser and Weinfurtner (SVW) formulated the most general HL theory with
projectability condition but without the detailed balance \cite{SVW}. Writing the 4-dimensional metric
in  the  ADM form,
 \bqn
 \lb{1.2}
ds^{2} &=& - N^{2}c^{2}dt^{2} + g_{ij}\left(dx^{i} + N^{i}dt\right)
     \left(dx^{j} + N^{j}dt\right), \nb\\
     & & ~~~~~~~~~~~~~~~~~~~~~~~~~~~~~~  (i, \; j = 1, 2, 3),~~~
 \eqn
the projectability condition requires that
\bq
\lb{1.4}
N = N(t), \;\;\; N^{i} = N^{i}(t, x),\;\;\; g_{ij} = g_{ij}(t, x).
\eq
Note that in \cite{WM,WWM,GPW}, the constant $c$, representing  the speed of light,  was absorbed into $N$.
The ADM form (\ref{1.2}) is preserved  by  the types of coordinate transformations,
\bq
\lb{1.3}
t \rightarrow f(t),\; \;\; x^{i} \rightarrow \zeta^{i}(t, {\bf x}).
\eq
Due to these restricted diffeomorphisms, one more degree of freedom appears
in the gravitational sector - a spin-0 graviton. This is potentially dangerous, and needs to decouple  
 in the Infrared (IR), in order to be consistent with observations. Similar problems are also found
in other modified theories, such as the massive gravity \cite{RT}.

Then, it can be shown that the most general action, which  preserves the parity, 
is given by \cite{SVW},
 \bqn \lb{1.6}
S = \zeta^2\int dt d^{3}x N \sqrt{g} \left({\cal{L}}_{K} -
{\cal{L}}_{{V}}+\zeta^{-2} {\cal{L}}_{M} \right),
 \eqn
where $g={\rm det}\,g_{ij}$, $ {\cal{L}}_{M}$ denotes the matter Lagrangian density,
and
 \bqn \lb{1.7}
{\cal{L}}_{K} &=& K_{ij}K^{ij} - \left(1-\xi\right)  K^{2},\nb\\
{\cal{L}}_{{V}} &=& 2\Lambda - R + \frac{1}{\zeta^{2}}
\left(g_{2}R^{2} +  g_{3}  R_{ij}R^{ij}\right)\nb\\
& & + \frac{1}{\zeta^{4}} \left(g_{4}R^{3} +  g_{5}  R\;
R_{ij}R^{ij}
+   g_{6}  R^{i}_{j} R^{j}_{k} R^{k}_{i} \right)\nb\\
& & + \frac{1}{\zeta^{4}} \left[g_{7}R\nabla^{2}R +  g_{8}
\left(\nabla_{i}R_{jk}\right)
\left(\nabla^{i}R^{jk}\right)\right],
 \eqn
where $\zeta^{2} = 1/{16\pi G}$, and 
the covariant derivatives and
Ricci and Riemann terms are all constructed from the three-metric $g_{ij}$,
while $K_{ij}$ is the extrinsic curvature,
 \bq \lb{1.8}
K_{ij} = \frac{1}{2N}\left(- \dot{g}_{ij} + \nabla_{i}N_{j} +
\nabla_{j}N_{i}\right),
 \eq
where $N_{i} = g_{ij}N^{j}$. 
In the IR limit, all the high order curvature terms (with
coefficients $g_{i},\; i = 2, ..., 8$) drop out, and the total action reduces when
$\xi = 0$ to the Einstein-Hilbert action.

\subsection{Linear Perturbations in de Sitter Background}

With the conformal time $\eta$, the de Sitter spacetime is given by
 $ ds^{2} = a^{2}(\eta)\left(- d\eta^{2} + \delta_{ij}dx^{i}dx^{j}\right)$,
 where $a(\eta) = - 1/(H\eta) = e^{Ht}$, 
 and $t$  denotes the cosmic time. 
  
Linear scalar perturbations of the metric are given by
 \bqn \lb{4.1}
\delta{g}_{ij} &=&   a^{2}(\eta)\left(- 2\psi\delta_{ij} + 2E_{,ij}\right),\nb\\
\delta{N}_{i} &=&   a^{2}(\eta) B_{,i}   ~~ \delta{N} = a(\eta) \phi(\eta).
 \eqn
Choosing the quasi-longitudinal gauge \cite{WM},
\bq
\lb{4.2}
\phi = 0 = E,
\eq
we find that the two gauge-invariant quantities defined in \cite{WM} reduce to,
\bq
\lb{4.3}
\Phi = {\cal{H}}B + B',\;\;\;
\Psi = \psi - {\cal{H}}B,
\eq
where  ${\cal{H}} = a'/a = -1/\eta$, and  $\psi$ and $B$ are given by \cite{HWW}
  \bqn
  \lb{4.8a}
& & \left(2 - 3\xi\right){\psi_{k}}'  = - {\xi}k^{2} B,\\
  \lb{4.8b}
& & {\psi_{k}}'' + 2{\cal H}{\psi_{k}}' + \omega^{2}_{k}\psi_{k} = 0,  
 \eqn
 in the momentum space, where
 \bq
 \lb{4.9}
 \omega^{2}_{k} = \left|c_{\psi}\right|^{2}k^{2} \left(-1 + \frac{k^{2}}{M_{A}^{2} a^{2}}
 +  \frac{k^{4}}{M_{B}^{4} a^{4}}\right),
 \eq
 with $ c^{2}_{\psi} \equiv {\xi}/{(2 - 3\xi)}$ and
 \bqn
 \lb{4.10}
 M_{A} &\equiv& \frac{M_{pl}}{\left[2(8g_{2} + 3g_{3})\right]^{1/2}},\nb\\
  M_{B} &\equiv& \frac{M_{pl}}{\left[4(8g_{7} - 3g_{8})\right]^{1/4}}.
 \eqn
 Clearly, to have $M_{A}$ and $M_{B}$ real, we must  assume  that 
 \bq
 \lb{4.10a}
 8g_{2} + 3g_{3} \ge 0,\;\;\; 8g_{7} - 3g_{8} \ge 0,
 \eq
conditions we shall take for granted in  the rest of this paper. 
 Note that in writing the above expressions, we had assumed that $\xi \le 0$. When $\xi = 0$ the corresponding solutions
 are stable, as shown in \cite{HWW}, so in the following we shall not consider this case any further, and concentrate
 ourselves only to the case $\xi < 0$. Then,  from the above one can see
 that the studies of stability of the de Sitter spacetime reduces to the studies of the master equation (\ref{4.8b}). Once
 $\psi_{k}$ is known, from Eq. (\ref{4.8a}) one can find $B_{k}$. Then, the gauge-invariant quantities 
 $\Phi_{k}$ and $\Psi_{k}$ can be read off from Eq.(\ref{4.3}). From the latter one can see that the properties
 of $\Phi_{k}$ and $\Psi_{k}$ are uniquely determined by $\psi_{k}$. In particular, if $\psi_{k}$ is  not singular, so
 are $\Phi_{k}$ and $\Psi_{k}$. Therefore, in the following we shall concentrate ourselves only on $\psi_{k}$.
 
 To study the perturbations further, we notice that Eq.(\ref{4.8b}) is quite similar to an oscillator with a dissipative force
 $ {\cal{F}}$ \cite{Gold}, 
\bq
\lb{4.11a}
 \ddot{x} + {\cal{F}}\dot{x} + \omega^{2} x = 0,
\eq
which   has the general solution,
\bq
\lb{4.11b}
x = A e^{-  {\cal{F}} t/2} e^{-i \omega t},
\eq
where $A$ is a constant. When ${\cal{F}} > 0$, from the above expression we can see that the free modes $\omega$
is exponentially damped. 

In the Minkowski background, we have $a =$ Constant. Without loss of generality, we can set $a = 1$. Then, we find
that  ${\cal{F}} = 0$, and 
\bq
\lb{4.11c}
\omega^{2}_{k} = - \left|c_{\psi}\right|^{2}k^{2} \left(1 - \frac{k^{2}}{M_{A}^{2}}
 -  \frac{k^{4}}{M_{B}^{4}}\right),\;\; (a = 1).
\eq
Therefore, if the scale of a mode is large enough so that $\omega^{2}_{k}$ becomes negative, this mode is unstable.
In particular, without the high-order corrections, all the modes are unstable  \cite{WM}. This is quite similar to the Jeans 
instability \cite{Wein}, for which there exists a characteristic  Jeans length $\lambda_{J} = 1/k_{J}$, where when scales
are smaller than the Jeans length, the modes are  stable.  When scales are larger than the Jeans length, they become 
unstable.  The largest instability occurs at 
\bq
\lb{4.11ca}
k^{2}_{M} = \frac{M_{B}^{2}}{\sqrt{r^{4} + 3} + r^{2}}, 
\eq
for which we have
\bqn
\lb{4.11cb}
\omega_{k}(k_{M}) &=& i \frac{\left|c_{\psi}\right|M_{B}}{{\cal{B}}^{3/2}}\left(r^{4} + r^{2}\sqrt{r^{4} + 3} + 2\right)^{1/2}
\nb\\
&\equiv& i\Gamma,
\eqn
where
\bqn
\lb{4.11cd}
{\cal{B}} &\equiv & \sqrt{r^{4} + 3} + r^{2},\nb\\
r &\equiv& \frac{M_{B}}{M_{A}} = \left(\frac{\left(8g_{2} + 3g_{3}\right)^{2}}{8g_{7} - 3g_{8}}\right)^{1/4}.
\eqn
The instability will grow significantly during a time $ t \ge t_{\Gamma} \equiv \Gamma^{-1}$, or in other words,   for any given time 
$t_{0}$ of interest, only when $t_{0} < t_{\Gamma}$, the growth of the instability during $t_{0}$ can be neglected. 
 
\begin{figure}[tbp]
\centering
\includegraphics[width=8cm]{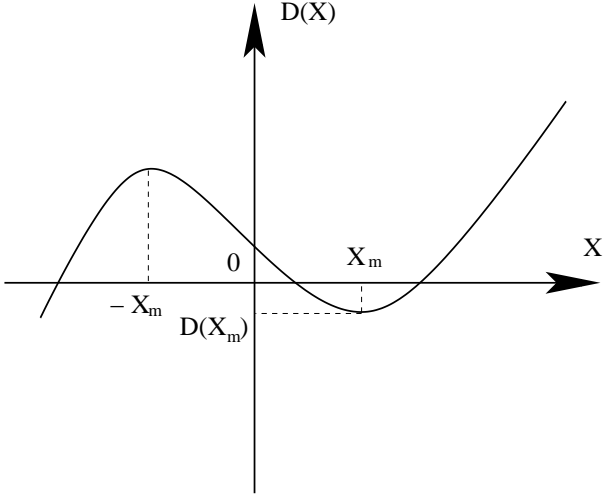}
\caption{The function $D(X)$ defined by Eq.(\ref{4.11i}). Note that only the half plane $X > 0$ is valid,
as one can see from Eq.(\ref{4.11f}). } 
\label{fD}
\end{figure}

However, it is well-known that Jeans instability can be removed by Hubble friction in an expanding universe \cite{Wein}. In the following 
we shall show that this is also true in the HL theory.   In particular, in the de Sitter background,   two important modifications occur:  
(a) For any given $k$,   $\omega_{k}^{2}$ is always positive at sufficiently early time, due to high-order corrections, as one can see  
from Eq.(\ref{4.9}). (b) The damping force ${\cal{F}} \; [= - 2/\eta]$ is strictly non-negative and independent of $H$. 
When  $\eta \rightarrow 0^{-}$ it becomes infinitely large. For short-scale waves, although the spacetime can be considered as 
locally flat, the high-order derivatives can kick in at a very early time, if the UV cutoff scale is very low. As time increases, the 
damping force becomes more and more important, and will finally become dominant.  Therefore, if the UV cutoff is sufficiently low, 
by combining these two kinds of effects, one works in the IR  ($\eta \simeq 0^{-}$) and the other works in the UV  ($|\eta| \gg 1$),    
one might be able to  stabilize the modes of both the short- and large-scales.   To see that this is indeed possible here in the HL theory, 
we first notice that, as the universe expands, $a$ becomes larger and larger, and there exists a moment, say,  $\eta_{c}$,  
at which $\omega_{k}^{2}(\eta_{c}) = 0$, where
\bq
\lb{4.11d}
\eta_{c}(k) = - \frac{\sqrt{2} M_{B}}{H k \left(r^{2} + \sqrt{r^{4} + 4}\right)^{1/2}}.
\eq
From  this moment on, the instability starts to develop until $\eta = 0^{-}$, at which we have 
$\omega^{2}_{k}(0^{-}) = -  \left|c_{\psi}\right|^{2}k^{2}$. Note that for the modes with $k \gtrsim k_{stable}$, we have
$\left|\eta_{c}(H_{0})/\eta_{0}\right| \lesssim 1$, that is,  the instability of these modes has
not occurred within the age of our universe, where $\eta_{0} \simeq -H^{-1}_{0}$ denotes the current conformal time of our universe, and 
\bq
\lb{4.11dd}
k_{stable} \equiv   \left(r^{2} + \sqrt{r^{4} + 4}\right)^{1/2} M_{B}.
\eq
Therefore, the only possible unstable modes that occur within the age of our universe are those with their wavelengths 
$\lambda > \lambda_{stable}$, where   $\lambda_{stable} \equiv  1/k_{stable}$. 
However, if   the UV cutoff scale $M_{*}$   is low enough, so that
the exponentially damping force kicks in before these   modes become unstable, 
that is, if
\bq
\lb{4.11e}
 {\cal{F}}(\eta) - 2 \left|\omega_{k}(\eta)\right| \ge 0,\; (\eta > \eta_{c}),
 \eq
then these unstable modes will be stabilized, and never show up, where
\bq
\lb{4.11ee}
M_{*} \; = min \{M_{A},\; M_{B}\}.
\eq  
Setting
\bq
\lb{4.11f}
X \equiv H^{2}\eta^{2} + \frac{M^{4}_{B}}{3k^{2}M^{2}_{A}},
\eq
the condition (\ref{4.11e}) can be written as
\bq
\lb{4.11g}
{{D}}(X) \equiv X^{3} - 3bX + 2d \ge 0,
\eq
where
\bqn
\lb{4.11h}
b &\equiv& \frac{M^{4}_{A}r^{4}}{9k^{4}}\Big(r^{4} + 3\Big),\nb\\
d &\equiv& \frac{M^{6}_{B}}{27k^{6}}\left(r^{6} + \frac{9}{2}r^{2} + \frac{27H^{2}}{2\left|c_{\psi}\right|^{2}M^{2}_{B}}\right).
\eqn
Fig. \ref{fD} schematically shows the function $D(X)$, from which we can see that the condition
(\ref{4.11g}) holds when 
\bqn
\lb{4.11i}
D\left(X_{m}\right) &=& \frac{2M^{12}_{B}}{3^{6}\left(d + b^{3/2}\right)k^{12}}\left[- \left(r^{4} + 3\right)^{3}\right.\nb\\
& & \left. + \left(r^{6} + \frac{9}{2}r^{2} 
+ \frac{27H^{2}}{2\left|c_{\psi}\right|^{2}M^{2}_{B}}\right)^{2}\right] \ge 0, ~~~~~~
\eqn
where $X_{m} = \sqrt{b}$. This yields
\bq
\lb{4.11j}
 M_{B}  \le   \Lambda_{stable}, 
\eq
where
\bq
\lb{4.11k}
 \Lambda_{stable} \equiv  \frac{H}{\left|c_{\psi}\right|}   \left\{\frac{2}{r^{4} + 4}\Bigg[\Big(r^4 + 3\Big)^{3/2} + r^2\Big(r^4 + \frac{9}{2}\Big)\Bigg]\right\}^{1/2}.
\eq
It is remarkable to note that the condition (\ref{4.11j}) does not depend on $k$. As a result, it is valid for any scale of modes.
In particular, once it is satisfied,  the short-scale modes become stabilized, too.    
Thus, for any given $M_{A}$ and $M_{B}$, if $\xi$ is sufficiently closed to its fixed point
$\xi= 0$ (at which one has $c_{\psi} = 0$)  \footnote{It should be noted that $c_{\psi}$ cannot be too closed to zero. Otherwise, Cherenkov radiation will impose   severe constraints.
We thank Thomas Sotiriou for pointing out this to us.  }, $ \Lambda_{stable}$ becomes large, and  the condition (\ref{4.11j}) can be  easily  satisfied. 
At the fixed point $\xi = 0$, we have   $ \Lambda_{stable} = \infty$, that is, now for any given $g_{i}$ (or equivalently for any
given $M_{A}$ and $M_{B}$), all the modes, of large- and small-scales, are stable.

The above can be further seen from the following limiting cases. First, when $r \ll 1$,  we have
\bq
\lb{4.11l}
 \Lambda_{stable}= \left(\frac{27}{4}\right)^{1/4}\frac{H}{\left|c_{\psi}\right|}, \; (r \ll 1).
\eq
Thus, even  $H$ is taken to be the current Hubble constant $H_{0}$,   $\Lambda_{stable}$ can still be  large, as longer as
$c_{\psi}$ is sufficiently closed to  its fixed point $c_{\psi} = 0$.  

When $r \simeq 1$,  on the other hand, we have
\bq
\lb{4.11m}
\Lambda_{stable} = \sqrt{\frac{27}{5}}\frac{H}{\left|c_{\psi}\right|}, \; (r \simeq 1).
\eq
Once again, if $\xi$ is sufficiently closed to    $\xi = 0$, $\Lambda_{stable}$ will be  large,
and the condition $M_{B} \le \Lambda_{stable}$ can be satisfied for a given non-zero $H$. 

When $r \gg 1$,  Eq.(\ref{4.11j}) reduces to
\bq
\lb{4.11o}
M_{A} \le   \frac{2H}{\left|c_{\psi}\right|}, \; (r \gg 1).
\eq

 Taking $H = H_{0}$, 
one can see that the conditions (\ref{4.11l})-(\ref{4.11o}) can be written as
\bq
 \lb{age}
 M_{*} \lesssim {\cal{O}}(1) \frac{H_{0}}{\left|c_{\psi}\right|},
 \eq
 which is equivalent to the condition that the instability found in the Minkowski background does not happen within the age of our
 universe \cite{KA,BPSc}.

\begin{figure}[tbp]
\centering
\includegraphics[width=8cm]{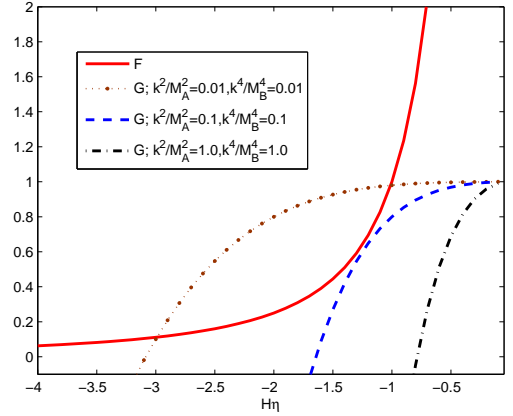}
\caption{The functions $F \equiv  {\cal{F}}^{2}/(4H^{2})$ and $G \equiv - \omega_{k}^{2}/H^{2}$   for
different choices of $k^{2}/M^{2}_{A}$ and $ k^{4}/M^{4}_{B}$, where 
 $\omega_0 \equiv \left|c_{\psi}\right|k/H$ has been sent to one in all the cases. When $k^{2}/M^{2}_{A}$ and $ k^{4}/M^{4}_{B}$
 are  small, $F- G = 0$ has two solutions $\eta_{1}$ and $\eta_{2}$ where $\eta_{c} < \eta_{1} < \eta_{2} < 0$, as shown by the
 doted line.  When $k^{2}/M^{2}_{A}$ and $ k^{4}/M^{4}_{B}$
 are large, $F$ is always greater than $G$, and $F - G = 0$ has no solution, as shown by the dash-dot line. 
 The dashed line represents  the case where $F - G = 0$
 has only one solution.}
\label{fg}
\end{figure}

Note that,  since the damping force always dominates when $\left|\eta\right| \ll  0$, 
 even the instability develops, it always occur within the period, $ \eta_{1} < \eta < \eta_{2}$, and will  be 
 finally  stabilized  by   ${\cal{F}}$, where $\eta_{c} < \eta_{1} \le \eta_{2} < 0$ [as can be seen from Fig. \ref{fg}],
 where $\eta_{1,2}$ are the two real and positive roots of $D(X) = 0$. Thus, for a given  time interval of interest,
 if the instability happens in a sufficient short period and will not grow much, before the damping force takes
 over, then such an instability is still acceptable. However, the following results show that this is possible only for
 modes of short-scales. In fact,   it can be shown that
  \bqn
 \lb{4.11p}
 \Delta\eta &\equiv&  \eta_{2} - \eta_{1} \nb\\
 &=&  \frac{2M_{A}^{2} r^{2}}{k^{2}}\sqrt{\frac{r^{4} + 3}{3}} \; \cos\left(\frac{2\theta + \pi}{6}\right), 
 \eqn
 where $\theta \in [\pi/2,\; \pi]$ and is defined as
 \bq
 \lb{4.11q}
 \cos\theta = - \frac{1}{\big(r^{4} + 3\big)^{3/2}}\left(r^{6} + \frac{9}{2}r^{2} 
 + \frac{27H^{2}}{2\left|c_{\psi}\right|^{2}M^{2}_{B}}\right).
 \eq
 Clearly, to have a real $\theta$, the denominator of Eq.(\ref{4.11q}) has to be greater or at least equal to the nominator,
 which is equivalent to  $M_{B} > \Lambda_{stable}$, where $\Lambda_{stable}$ is defined in Eq.(\ref{4.11j}).
 Since $\Delta\eta \propto k^{-2}$, one can see that, for any given $\xi, g_{i}$ and $H$,  $\Delta\eta \rightarrow \infty$
 when $ k^{2} \rightarrow 0$. Thus, to limit the instability completely for any $k$, one needs to
 require that the condition Eq.(\ref{4.11j}) hold strictly. 

\begin{figure}[tbp]
\centering
\includegraphics[width=8cm]{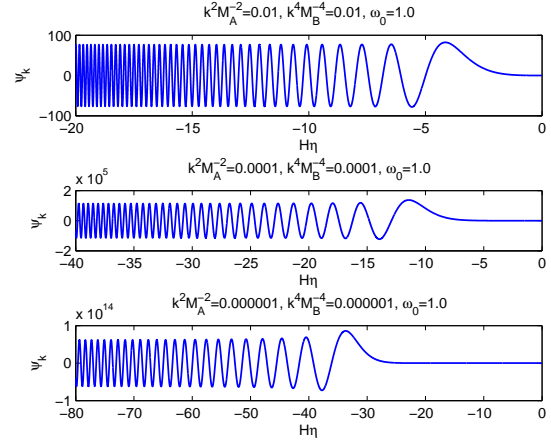}
\caption{The metric perturbation $\psi_{k}$, in the cases where both  $k^{2}/M^{2}_{A}$ and $ k^{4}/M^{4}_{B}$ are very small, so that
 $F- G = 0$ has two solutions $\eta_{1}$ and $\eta_{2}$ where $\eta_{c} < \eta_{1} < \eta_{2} < 0$, as shown in Fig. \ref{fg}. It can be seen 
 that $\psi_{k}$ oscillates with almost a constant amplitude when $H\eta \ll 0$, and then  increases a bit, before it decays rapidly 
  to zero, where we had chosen $c_1 = c_2$.}
\label{fig0}
\end{figure}

 Fig. \ref{fig0} shows the case where  $M_{B} > \Lambda_{stable}$ with a finite and non-zero $k$, from which we can see that 
 the mode is oscillating all the way down to $H\eta = - {\cal{O}}(10)$,
 and then grows a little bit, before it starts to decay. Since the decaying rate is very large (inversely proportional to $-\eta$,
 as one can see from Eq.(\ref{4.11b}) where ${\cal{F}} = - 2/\eta$),
  it dies away rapidly afterwards.   
 
\begin{figure}[tbp]
\centering
\includegraphics[width=8cm]{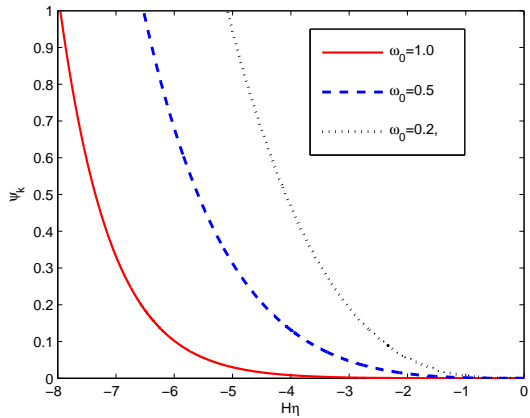}
\caption{The metric perturbation $\psi_{k}$, defined by Eq.(\ref{4.1}) in the quasi-longitudinal gauge for
different choices of $\omega_0$, with $k^{2}/M^{2}_{A}  = 0$ and $ k^{4}/M^{4}_{B}  = 0$.
In all the three cases we have set $c_1 = c_2$. }
\label{fig1}
\end{figure}

Our above analytical analysis is further supported by the following numerical calculations. Let us first notice that 
$\omega^{2} \rightarrow - \left|c_{\psi}\right|^{2}k^{2}$ as $H\eta \rightarrow 0^{-}$.
Then,  the asymptotical solution of Eq.(\ref{4.8b}) satisfies the equation,
\bq
\lb{4.12}
 {\psi_{k}}'' - \frac{2}{\eta} {\psi_{k}}' - \left|c_{\psi}\right|^{2}k^{2}\psi_{k} = 0,
 \eq
which has the general solution  \cite{HWW}, 
\bqn
\lb{4.13}
\psi_{k} &=& {c}_{1}\big(z - 1\big)e^{z} +  {c}_{2}\big(z+1\big)e^{-z},\nb\\
B_{k} &=& \frac{(3\xi -2)z}{\xi k^{2}} \Big({c}_{1} e^{z} -  {c}_{2} e^{-z}\Big),
\eqn
where $z \equiv \left|c_{\psi}\right| k \eta = -(\left|c_{\psi}\right| k/H)e^{-Ht}$. 
Clearly, they are all finite as $H\eta \rightarrow 0^{-}$ (or $t \rightarrow \infty$). In particular, $B_{k} \rightarrow 0$ and $\psi_{k} \rightarrow 
c_2 - c_1 $ in the IR limit $H\eta \rightarrow 0^{-}$. Note the slit difference between the  
 two constants $c_{1,2}$ defined here and the ones used in \cite{HWW}. Fig. \ref{fig1} shows the function  $\psi_{k}(\eta)$ 
 with different   choices of $\omega_{0} [\equiv \left|c_{\psi}\right|k/H]$,  from which we can see that 
  the larger $\omega_{0}$ is, the faster $\psi_{k}$ decays.  That is,   small-scale modes always decay faster
 than large-scale ones. 
 
\begin{figure}[tbp]
\centering
\includegraphics[width=8cm]{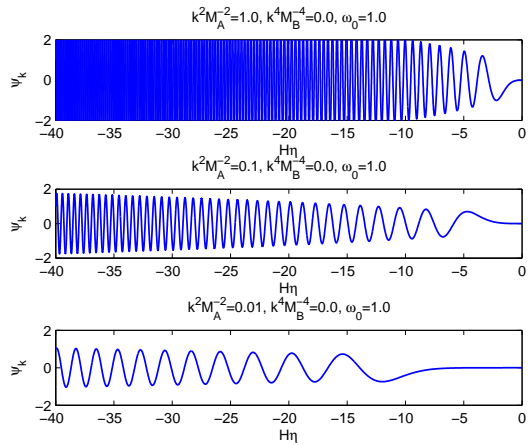}
\caption{The function $\psi_{k}$, defined by Eq.(\ref{4.1}) 
for different choices of  $k^{2}/M^{2}_{A}$ with $\omega_0   = 1$ and $ k^{4}/M^{4}_{B}  = 0$.}
\label{fig2}
\end{figure}

\begin{figure}[tbp]
\centering
\includegraphics[width=8cm]{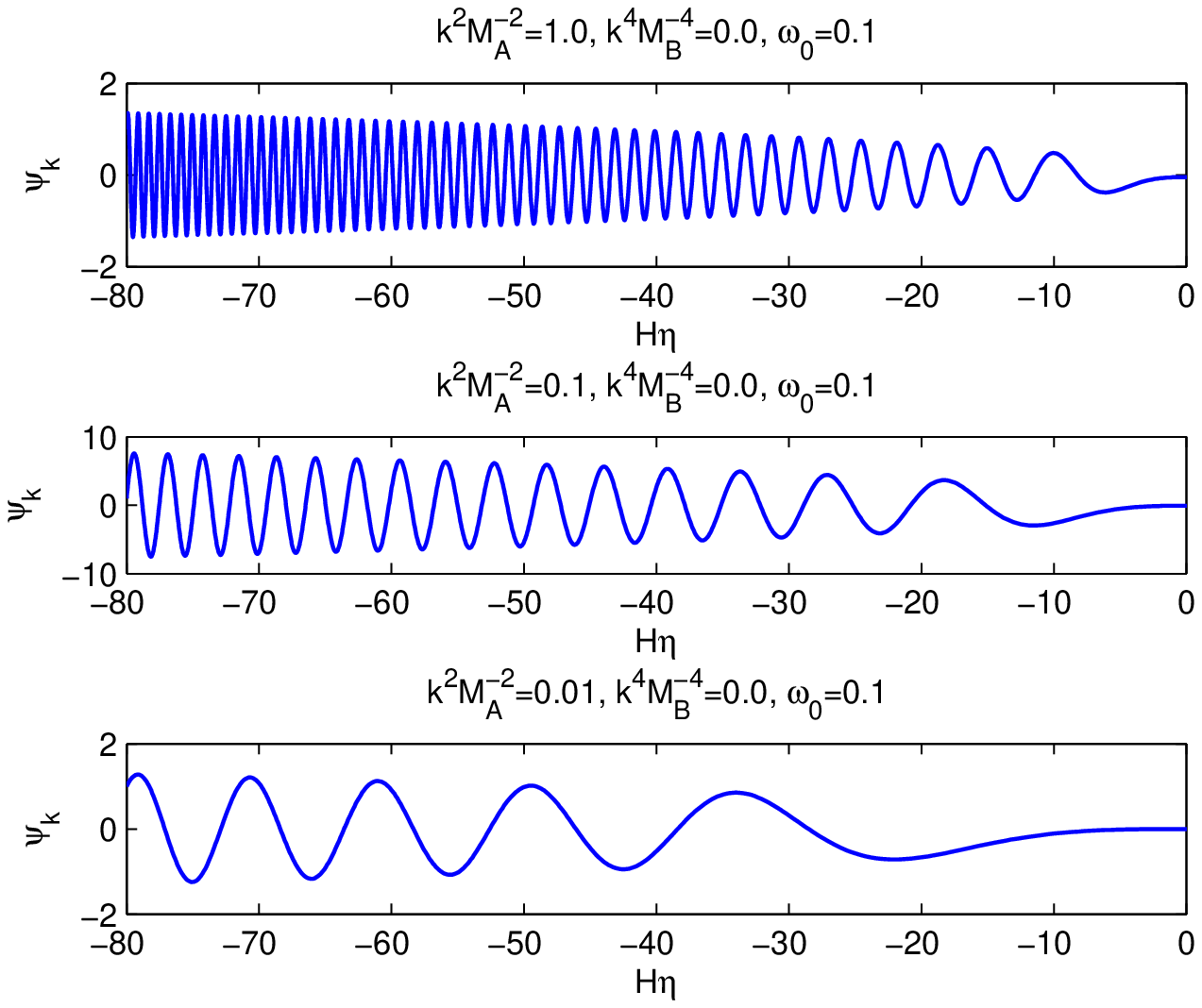}
\caption{The function $\psi_{k}$, defined by Eq.(\ref{4.1}) 
for different choices of  $k^{2}/M^{2}_{A}$ with $\omega_0   = 0.1$ and $ k^{4}/M^{4}_{B}  = 0$.}
\label{fig3}
\end{figure}

\begin{figure}[tbp]
\centering
\includegraphics[width=8cm]{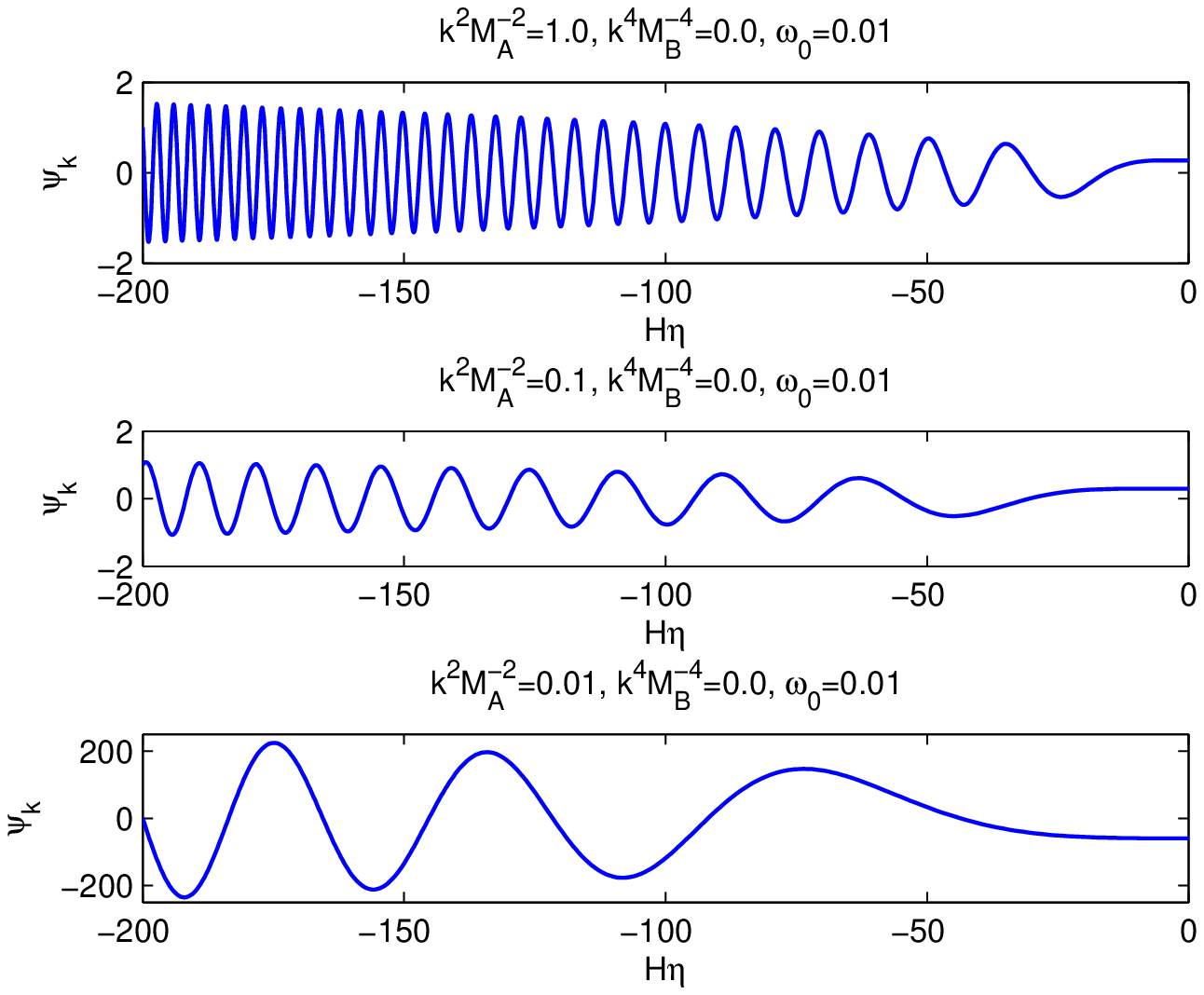}
\caption{The function $\psi_{k}$, defined by Eq.(\ref{4.1}) 
for different choices of  $k^{2}/M^{2}_{A}$ with $\omega_0   = 0.01$ and $ k^{4}/M^{4}_{B}  = 0$.}
\label{fig4}
\end{figure}

\begin{figure}[tbp]
\centering
\includegraphics[width=8cm]{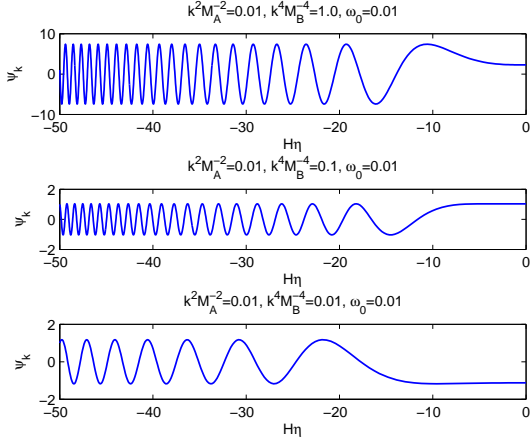}
\caption{The function $\psi_{k}$, defined by Eq.(\ref{4.1}) 
for different choices of  $k^{4}/M^{4}_{B}$ with $\omega_0   = 0.01$ and $ k^{2}/M^{2}_{A}  = 0.01$.}
\label{fig5}
\end{figure}

To study the effects of the high-order curvatures, we gradually turn on the  fourth- and sixth- order corrections. In particular,
Fig. \ref{fig2} shows the case where $\omega_0 \equiv \left|c_{\psi}\right|k/H = 1$ and  $k^{4}/M^{4}_{B} = 0$, with three different 
values of the suppressed scale, $M_{A}$. From there one can see that, when $M_{A}$ is small, the perturbation oscillates many times
before it starts to decay.  As $M_{A}$ increases, the oscillating times becomes less and less. 
The same characteristics  persist even for small values of $\omega_0$, as
shown by Figs. \ref{fig3} and \ref{fig4}. The effects of the sixth-order term $k^{4}/M^{4}_{B}$ are shown in Fig. \ref{fig5}.

In all the cases,   the perturbations will finally decay exponentially for any given $k$, as the damping force ${\cal{F}}$ 
is independent of $k$ and  ${\cal{F}} + 2i \omega_{k}(\eta) \simeq {\cal{F}} - 2 \omega_{0} \gg 1$  as $\eta \rightarrow 0^{-}$,
since   ${\cal{F}}(0^{-}) = \infty$. Therefore,  for any given $k$ the perturbations always decay exponentially 
as $\eta \simeq 0^{-}$ or ($t \rightarrow \infty$).   

\section{Strong Coupling }
\renewcommand{\theequation}{3.\arabic{equation}} \setcounter{equation}{0}

To understand the strong coupling problem,  we shall restrict ourselves mainly  to the  perturbations in
the Minkowski background, because such obtained results   can be easily generalized to the de Sitter background 
\cite{KA}.  Such a study also helps us to clarify some differences regarding to the strength of strong couplings, 
obtained recently  in \cite{KA,PS,BPSc,KP}.  In addition, the  treatment in this section can be applied to the HL 
theory both with and without the projectability condition. So, the conclusions obtained in this section are applicable 
to both cases. 
 
When the background is Minkowski,  without loss of generality, we consider the metric perturbations \cite{PS},
\bq 
\lb{3.1}
N = 1, \;\;\; {g}_{ij} = e^{2\zeta(t,x)} \delta_{ij}, \;\;\; {N}_{i} =   \partial_{i}\beta(t,x), 
\eq
which lead,  respectively,  to the following second- and third-order actions,
\bqn
\lb{3.2a}
S^{(2)} &=& M_{pl}^{2} \int{dt d^{3}x \left(- \frac{\dot{\zeta}^{2}}{c_{\psi}^{2}} + \big(\partial\zeta\big)^{2}\right)},\\
\lb{3.2b}
S^{(3)} &=& M_{pl}^{2} \int{dt d^{3}x \Bigg\{\zeta\big(\partial\zeta\big)^{2} - \frac{3\left(2c_{\psi}^{2} + 1\right)}{2c^{4}_{\psi}} \zeta {\dot{\zeta}}^{2} }\nb\\
& & 
- \frac{2}{c^{4}_{\psi}} \dot{\zeta}\partial_{i}\zeta \frac{\partial^{i}}{\Delta} \dot{\zeta}
+ \frac{3\zeta}{c^{4}_{\psi}} \left(\frac{\partial_{i}\partial_{j}}{\Delta}\dot{\zeta}\right)^{2}\Bigg\},
\eqn
where $\Delta \equiv \partial^{i}\partial_{i}$, and 
\bq
\lb{3.2c}
\Delta \beta = - \frac{1}{c^{2}_{\psi}}\dot{\zeta}.
\eq
Note that the above expressions can be easily obtained from the limit $\eta \rightarrow \infty$, where $\eta$ is the coupling constant of the gradient
term, $a_{i}a^{i}$, introduced in \cite{BPSb}, which should not be confused with the conformal time, used in the last section. In addition, in the
Minkowski background the conformal time is identical to   the cosmic time $t$. Then, comparing Eqs.(\ref{3.1})  and (\ref{4.1}), we find that $\zeta = 
-\psi$ and $\beta = a^{2}B$ to the linear order of perturbations for $a = 1$. 

To consider the strong coupling problem, we first write the  quadratic action $S^{(2)}$ in a canonical form with unity coupling constants. This can be done by 
the coordinate  transformations,
\bq
\lb{3.4}
t =\alpha \hat{t},\;\;\; x^{i} = \beta \hat{x}^{i},
\eq
which are allowed by the gauge freedom (\ref{1.3}), where  $\alpha$ and $\beta$ are arbitrary constants. Choosing
\bq
\lb{3.5}
\zeta =  \frac{\hat{\zeta}}{M_{pl}\left|c_{\psi}\right|^{1/2} \alpha}, \;\;\; \beta = \left|c_{\psi}\right| \alpha, 
\eq
one finds that $S^{(2)}$ given by Eq.(\ref{3.2a}) can be   written as,  
\bq
\lb{3.6}
S^{(2)}=   \int{d\hat{t} d^{3}\hat{x} \Big(\hat{\zeta}^{*2}   + \big(\hat{\partial}\hat{\zeta}\big)^{2}\Big)},
\eq
where $\hat{\zeta}^{*} \equiv d\hat{\zeta}/d\hat{t}$, while the cubic action $S^{(3)}$ takes the form,
\bqn
\lb{3.8}
S^{(3)} &=& \frac{1}{\Lambda_{SC}} \int{d\hat{t} d^{3}\hat{x} \Bigg\{\frac{2\left|c_{\psi}\right|^{2}}{3} \hat{\zeta} \big(\hat{\partial}\hat{\zeta}\Big)^{2} 
+ \hat{\zeta}\left(\frac{\hat\partial_{i}\hat\partial_{j}}{\hat\Delta}\hat{\zeta}^{*}\right)^{2}}\nb\\
&& - \frac{4}{3} \hat{\zeta}^{*}\hat{\partial}_{i}\hat{\zeta}\frac{\hat{\partial}^{i}}{\hat{\Delta}}\hat{\zeta}^{*}
- \Big(1 - 2\left|c_{\psi}\right|^{2}\Big)\hat{\zeta}\hat{\zeta}^{*2}\Bigg\},
\eqn
where
\bq
\lb{3.9}
\Lambda_{SC} \equiv \frac{2}{3} M_{pl}\left|c_{\psi}\right|^{5/2}\alpha.
\eq
Clearly, if one chooses $\alpha \propto \left|c_{\psi}\right|^{-5/2}$, one finds that $\Lambda_{SC}$ will remain finite when $c_{\psi} \rightarrow 0$.
In the following we shall choose $\alpha = 3\left|c_{\psi}\right|^{-5/2}/2$, so that $\Lambda_{SC} = M_{pl}$. 

Requiring that the quadratic action $S^{(2)}$ be invariant under the  rescaling \cite{Pol},
\bq
\lb{3.10}
\hat{t} \rightarrow s^{-\gamma_{1}} \hat{t},\;\;\; \hat{x} \rightarrow s^{-\gamma_{2}} \hat{x},\;\; \hat{\zeta} \rightarrow s^{\gamma_{3}} \hat{\zeta},
\eq
we find that $\gamma_{1} = \gamma_{2} = \gamma_{3}$. Without loss of generality, we can always choose $\gamma_i  = 1 \;(i = 1, 2, 3)$, so that Eq.(\ref{3.10}) is identical to the 
relativistic scaling. Then, it can be shown that all the terms in the cubic action (\ref{3.8}) scale as $s^{1}$, which means that these terms are
irrelevant in the low energy limit, but diverge  in the UV, so they are not renormalizable \cite{Pol}. This indicates that the perturbations break 
down when the coupling coefficients greatly exceed units. To calculate these  coefficients, let us consider a process at the energy scale $E$, then we find that
all the terms in the cubic action has the same magnitude as $E$, for example,
\bq
\lb{3.11}
\int{d\hat{t} d^{3}\hat{x}   \hat{\zeta} \big(\hat{\partial}\hat{\zeta}\big)^{2}} \simeq E.
\eq
Since the action is dimensionless, all the  coefficients in (\ref{3.8}) must have the dimension $E^{-1}$. Writing them in the form,
\bq
\lb{3.12}
\lambda_{i} = \frac{\hat{\lambda}_{i}}{\Lambda_{i}},
\eq
where $\hat{\lambda}_{i}$ is a dimensionless parameter of order one, we find that the lowest scale of $\Lambda_{i}$'s is given by the last three terms in
Eq.(\ref{3.8}) and is of the order of $\Lambda_{SC}$. 
Translating it back to the coordinates $t$ and $x$, the corresponding
energy and momentum scales are,
\bqn
\lb{3.13a}
\Lambda_{\omega} &=& \frac{\Lambda_{SC}}{\alpha} \simeq \left|c_{\psi}\right|^{5/2}M_{pl},\nb\\
\Lambda_{k} &=& \frac{\Lambda_{SC}}{\beta} \simeq \left|c_{\psi}\right|^{3/2}M_{pl},
\eqn
which are consistent with the results obtained in \cite{BPSb,BPSc} by using the St\"uckelberg trick (See also \cite{KA}), 
but slightly different from that given in \cite{PS}. 

As $c_{\psi} \rightarrow 0$, these scales vanish, indicating that strong coupling happens when $c_{\psi}$ is very small.  
For processes with momentum $k \gtrsim \Lambda_{k}$,  the problem becomes
strong coupling, and non-linear  effects are important and must be taken into account. Mukohyama recently showed that these  
effects make the spin-0 graviton finally decoupled, and the relativistic limit $\xi \rightarrow 0$ in the IR exists for spherically symmetric, 
static, vacuum 
spacetimes \cite{Mukc}.  

 \begin{figure}[tbp]
\centering
\includegraphics[width=8cm]{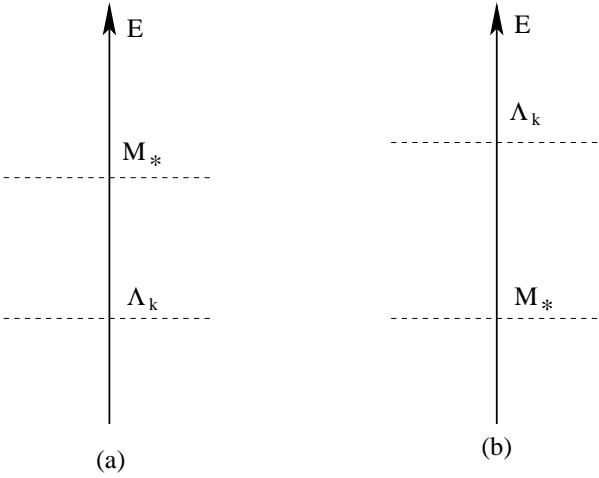}
\caption{The energy scales: (a) $\; \Lambda_{k} \lesssim M_{*}$; and (b) $\; \Lambda_{k} \gtrsim M_{*}$.}
\label{fig6}
\end{figure}

It must be noted that the above analysis is valid only for $M_{*} \gtrsim \Lambda_{k}$  [cf. Fig.\ref{fig6}(a)]. 
If 
\bq
\lb{3.14a}
M_{*} \lesssim \Lambda_{k}, 
\eq
which is the precise condition for the BPS mechanism to work [cf. Eq.(\ref{1.b})],
then the high-order derivative terms become important before the strong coupling energy scale $\Lambda_{k}$ reaches, 
and the above analysis is   no longer valid  [cf. Fig.\ref{fig6}(b)]. 
Including the high order derivatives, one finds that the quadratic 
 action becomes, 
 \bqn
\lb{3.14}
S^{(2)} &=& M_{pl}^{2} \int{dt d^{3}x \left(- \frac{\dot{\zeta}^{2}}{c_{\psi}^{2}} + \big(\partial\zeta\big)^{2} - \frac{1}{M_{A}^{2}}\zeta\partial^{4}\zeta\right.}\nb\\
& & ~~~~~~~~~~~~~~~~~~~~ {\left. + \frac{1}{M_{B}^{4}}\zeta\partial^{6}\zeta \right)}.
\eqn
Depending on whether   $M_{B}< M_{A}$ or $M_{B} > M_{A}$, the low energy behavior will be different. 
In the following, let us  consider them separately.

\subsection{$M_{B} < M_{A}$}

In this case,  we have $M_{*} = M_{B}$. Then, we can see that
the sixth-order derivative term will dominate the fourth-order one. If we consider a process at the momentum scale $k \gtrsim M_{B}$, then the first and last
terms in $S^{(2)}$ will be dominant. Using the coordinate transformation (\ref{3.4}) and the rescaling of $\zeta = \gamma \hat{\zeta}$, we first transform
these terms to the ones with unit coefficients. It can be shown that this can be realized by choosing
\bq
\lb{3.15}
\alpha = \frac{M_{B}}{\left|c_{\psi}\right|}\beta^{3}, \;\;\;
\gamma = \frac{M_{B}}{M_{pl}}\left|c_{\psi}\right|^{1/2},
\eq
where $\beta$ is arbitrary. Then, we obtain that 
 \bqn
\lb{3.16}
S^{(2)} &=&   \int{d\hat{t} d^{3}\hat{x} \left(\hat{\zeta}^{*2} + \beta^{4}M_{B}^{4} \big(\hat{\partial}\hat{\zeta}\big)^{2}
\right.}\nb\\
& & ~~ { - \frac{M_{B}^{4}\beta^{2}}{M_{A}^{2}}\hat{\zeta}\hat{\partial}^{4}\hat{\zeta}
 +  \hat{\zeta}\hat{\partial}^{6}\hat{\zeta}\Big)},
\eqn
and
\bqn
\lb{3.17}
S^{(3)} &=& \frac{1}{\Lambda_{SC}^{(B)}} \int{d\hat{t} d^{3}\hat{x} \Bigg\{
\beta^{4}M_{B}^{4} \hat{\zeta} \Big(\hat{\partial}\hat{\zeta}\Big)^{2} 
}\nb\\
&& - \frac{3}{2}\Big(1 - 2\left|c_{\psi}\right|^{2}\Big)\hat{\zeta}\hat{\zeta}^{*2} 
- 2 \hat{\zeta}^{*}\hat{\partial}_{i}\hat{\zeta}\frac{\hat{\partial}^{i}}{\hat{\Delta}}\hat{\zeta}^{*}\nb\\
& &
+ 3 \hat{\zeta}\left(\frac{\hat\partial_{i}\hat\partial_{j}}{\hat\Delta}\hat{\zeta}^{*}\right)^{2} 
+ ...\Bigg\},
\eqn
 with 
 \bq
 \lb{3.18}
 \Lambda_{SC}^{(B)} =  \frac{M_{pl}}{M_{B}}\left|c_{\psi}\right|^{3/2}. 
 \eq
The ``..."  in $S^{(3)}$ represents the cubic terms coming from the high-order derivative corrections, such as $f_{1}\zeta^{2}\partial^{4}\zeta$
and $f_{2}\zeta^{2}\partial^{4}\zeta$, where $f_{1}$ and $f_{2}$ are independent of $c_{\psi}$ and functions of the coupling 
constants $g_{i}$ only. As a result, the limit, $c_{\psi} \rightarrow 0$,  of these terms always finite, and have no contributions
to the strong coupling problem. In fact, it can be shown that these terms are either relevant or marginal (cf. the following analysis.). So, 
in the following, without loss of generality  we shall ignore them. 
%
Then,  under the re-scaling,
\bq
\lb{3.19}
t \rightarrow s^{-3} t,\;\;\; x  \rightarrow s^{-1} x, \;\;\; \zeta \rightarrow s^{0} \zeta,
\eq
the first and last terms in the right-hand side of Eq.(\ref{3.16}) are unchanged, while the second and third terms scale like
$s^{-4}$ and $s^{-2}$, respectively. Therefore, these terms are relevant and super-renormalizable. Similarly, the first term
in the cubic action $S^{(3)}$ of Eq.(\ref{3.17}) scales as $s^{-4}$, while all the rest scales as $s^{0}$, that is, the  term 
$\hat{\zeta} \big(\hat{\partial}\hat{\zeta}\big)^{2}$ is relevant, while the rest, the second, third and fourth in $S^{(3)}$,
are all marginal and strictly renormalizable. Thus, as the energy scale of the system changes, the amplitude of these latter
terms remain the same. That is, these terms are equally important at all scales of energy, provided that the condition (\ref{3.14a})
holds. Since they are all suppressed by the dimensionless quantity $\Lambda_{SC}^{(B)}$, we can see that in the present case 
the problem becomes strong coupling when $c_{\psi}$ is very small, unless $\Lambda_{SC}^{(B)} \gtrsim 1$, which is equivalent to,
\bq
\lb{3.19a}
M_{B} \lesssim M_{pl}\left|c_{\psi}\right|^{3/2}.
\eq

\subsection{$M_{B} > M_{A}$}

 When $M_{B} > M_{A}$, we have $M_{*} = M_{A}$. Then,  the fourth-order derivative term in the quadratic action $S^{(2)}$
 given by Eq.(\ref{3.14}) will dominate the sixth-order term. Then, the rescaling (\ref{3.4}) and $\zeta = \gamma \hat{\zeta}$ with
 \bq
 \lb{3.20}
 \alpha = \frac{1}{\left|c_{\psi}\right| M_{A}},\;\;\;
 \beta = \frac{1}{M_{A}},\;\;\;
 \gamma = \frac{M_{A}}{M_{pl}}\left|c_{\psi}\right|^{1/2},
 \eq
 will bring the quadratic action (\ref{3.14}) to the form,
  \bqn
\lb{3.21}
S^{(2)} &=&   \int{d\hat{t} d^{3}\hat{x} \Big(\hat{\zeta}^{*2} +   \big(\hat{\partial}\hat{\zeta}\big)^{2}}\nb\\
& & ~~ { -  \hat{\zeta}\hat{\partial}^{4}\hat{\zeta}
 + \left.\left(\frac{M_{A}}{M_{B}}\right)^{4}\hat{\zeta}\hat{\partial}^{6}\hat{\zeta}\right)},
\eqn
while the cubic action takes the form,
\bqn
\lb{3.22}
S^{(3)} &=& \frac{1}{\Lambda_{SC}^{(A)}} \int{d\hat{t} d^{3}\hat{x} \Bigg\{\left|c_{\psi}\right|^{2}
  \hat{\zeta} \Big(\hat{\partial}\hat{\zeta}\Big)^{2} 
}\nb\\
&& - \frac{3}{2}\Big(1 - 2\left|c_{\psi}\right|^{2}\Big)\hat{\zeta}\hat{\zeta}^{*2} 
- 2 \hat{\zeta}^{*}\hat{\partial}_{i}\hat{\zeta}\frac{\hat{\partial}^{i}}{\hat{\Delta}}\hat{\zeta}^{*}\nb\\
& &
+ 3 \hat{\zeta}\left(\frac{\hat\partial_{i}\hat\partial_{j}}{\hat\Delta}\hat{\zeta}^{*}\right)^{2} 
+ ...\Bigg\},
\eqn
where
\bq
\lb{3.23}
 \Lambda_{SC}^{(A)} \equiv  \frac{M_{pl}}{M_{A}}\left|c_{\psi}\right|^{3/2}. 
 \eq
 Similar to those given in Eq.(\ref{3.17}), the ``..."    are cubic terms of the forms $f_{1}(g_{s})\zeta^{2}\partial^{4}\zeta$
and $f_{2}(g_{s}) \zeta^{2}\partial^{4}\zeta$, which are all finite in the limit $\xi \rightarrow 0$, so they are irrelevant   
to the strong coupling problem.  

Then, we find that the first and third terms in the right-hand side of Eq.(\ref{3.21}) are unchanged, under the rescaling,
\bq
\lb{3.24}
t \rightarrow s^{-2} t,\;\;\; x  \rightarrow s^{-1} x, \;\;\; \zeta \rightarrow s^{1/2} \zeta,
\eq
for which the first term in the right-hand side of the cubic action $S^{(3)}$ given by Eq.(\ref{3.22}) scales as $s^{-3/2}$.
Thus, this term is relevant and super-renoralizable. The second, third and fourth terms, on the other hand, are all scale
as $s^{1/2}$, so they are all irrelevant and non-renormalizable. Then, if we consider processes at the energy scale
$E$, we find that $\int{d\hat{t} d^{3}\hat{x} \hat{\zeta}\hat{\zeta}^{*2}} \simeq E^{1/4}$, so that the second term in $S^{(3)}$
is suppressed by,
\bq
\lb{3.25}
\Lambda_{\hat{\omega}} = \left(\frac{M_{pl}}{M_{A}}\right)^{4}\left|c_{\psi}\right|^{6}.
\eq
It can be shown that the third and fourth terms are suppressed by the same factor. 
Transforming it back to the $(t, x^{i})$-coordinates, we find that
the energy and momentum are suppressed, respectively, by
\bqn
\lb{3.26}
\Lambda_{\omega} & = & \frac{\Lambda_{\hat{\omega}}}{\alpha}  =  \left(\frac{M_{pl}\left|c_{\psi}\right|^{7/4}}{M_{A}}\right)^{4} M_{A},\nb\\
\Lambda_{k} &=& \frac{\left(\Lambda_{\hat{\omega}}\right)^{1/2}}{\beta} =   \left(\frac{M_{pl}\left|c_{\psi}\right|^{3/2}}{M_{A}}\right)^{2} M_{A}.
\eqn
Then, the condition (\ref{3.14a}) implies that 
$M_{A} \lesssim M_{pl}\left|c_{\psi}\right|^{3/2}$, 
which, together with Eq.(\ref{3.19a}), can be written as
\bq
\lb{3.19b}
M_{*} \lesssim M_{pl}\left|c_{\psi}\right|^{3/2}.
\eq

If one takes the Minkowski spacetime as the legitimate background, as shown in \cite{SVW,WM}, it is not stable in the SVW setup,
and one would require that the instability should not show up within the age of the universe,
\bq
\lb{3.19d}
\left|c_{\psi}\right| \lesssim \frac{H_{0}}{M_{*}}.
\eq
BPS found that this, together with the condition (\ref{3.19b}), 
implies $M_{*} \lesssim (100 \; m)^{-1}$, or equivalent to
\bq
\lb{3.19da}
|\xi| \lesssim \left(\frac{H_{0}}{M_{*}}\right)^{2/5} \simeq 10^{-24}.
\eq
%
Clearly, this raises the fine-tuning problem, as a natural value of $\xi$ in the UV is expected to be order of one \cite{Horava}.
 It is unclear by which kind of mechanism it can be driven so closed to its
 relativistic value $\xi = 0$
   \cite{BPSc}  (See also \cite{KA}). 
 
Following \cite{KA}, it can be easily generalized the above studies to the de Sitter background, and similar conclusions will be
obtained: (a) When $M_{*} \gtrsim \Lambda_{k} = M_{pl}\left|c_{\psi}\right|^{3/2}$, the theory becomes
strong coupling for processes with energies $E \simeq   \Lambda_{k}$ [See  Eq.(\ref{3.13a})].   (b) When $M_{*} \lesssim \Lambda_{k}$,
the strong coupling problem does not exist. However, if one considers the studies of perturbations given in Sec. II as in the current 
universe, namely,  $H = H_{0}$, then the stability condition   is that of Eq.(\ref{age}),  which is the same as Eq.(\ref{3.19d}). Hence,
the results obtained above also apply to the de Sitter background with $H = H_{0}$. Therefore, it is 
concluded that {\em the mechanism, $M_{*} \le \Lambda_{k}$, of solving the   strong coupling problem
invented in \cite{BPSb,BPSc} for the HL theory without projectability condition,  cannot be applied to the case with projectability condition}.

Thus, in the SVW setup one  may choose the de Sitter spacetime as its legitimate background in order to avoid the instability problem.
In order to have a reasonable UV cutoff scale $M_{*}$, where now $M_{*}$  must satisfy the conditions,
\bq
\lb{3.19e}
M_{pl}\left|c_{\psi}\right|^{3/2} \lesssim M_{*} \lesssim  \frac{H_{0}}{\left|c_{\psi}\right|},
\eq
its IR limit  has to be very closed to, if not precisely at, the GR fixed point. Of course,   with such a choice, the theory is strong coupling. 
This will not be a problem, if the relativistic 
limit can be obtained after the non-linear effects are taken into account. In the spherically symmetric, static, vacuum spacetimes, 
Mukohyama showed that this is indeed the case \cite{Mukc}.  In the following section, we shall present a class of exact solutions of the theory,
from which one can show clearly  that the relativistic limit exists, and the limited spacetime is exactly the 
(rotating) de Sitter spacetime of GR.

\section{Non-Perturbative Cosmological Solutions}
\renewcommand{\theequation}{4.\arabic{equation}} \setcounter{equation}{0}

In the DGP model of branes \cite{DGP}, Newtonian approximations lead to a Friedmann equation with a constant $\tilde{G}$ that is different from
the  Newtonian constant $G$ by a factor $4/3$ \cite{Deffayet}, while the non-perturbative equation in the flat FRW universe with 
zero-cosmological constant takes the form,
$$
H^{2} = \frac{8\pi G}3 \rho - m_c H,
$$
 where $m_c$ is the graviton mass    \cite{DDGV}. Clearly, when $m_c \rightarrow 0$, it reduces precisely
 to the Friedmann equation in GR. This shows clearly that the spin-0 massive graviton decouples, when the non-linear 
 effects are taken into account, and, as a result, the theory smoothly passes over to the GR limit. 
 
 In this section, we shall 
 show that the same happens here in the SVW setup, too. That is, when we do the linear perturbations of the de Sitter 
 background, we have the strong coupling problem, as shown explicitly in the last section. But, the exactly solutions of the 
 theory have a smoothy GR limit. As a matter of fact, this can  already be seen clearly if one simply looks at the corresponding
 Friedmann equation \cite{HWW},
 \bq
 \lb{6.0}
 \left(1 - \frac{3}{2}\xi\right) H^{2} = \frac{8\pi G}3 \rho + \frac{1}{3}\Lambda.
 \eq
(The other independent equation is the well-known conservation law of energy and momentum, $\dot{\rho} + 3H(\rho + p) = 0$.
For detail, see \cite{SVW,WM,WWM}.) From the above expression we can see that replacing $G$ in all the solutions obtained in GR 
by $\tilde{G} \equiv G/(1- 3\xi/2)$, we shall obtain all the cosmological (flat) solutions in the HL theory. 

In the following, we shall go a little bit further, and show that this is true also in the sense of non-linear perturbations of the 
de Sitter spacetime. Let us first note that, once nonlinear effects are taken into account, the separation of scalar, vector and tensor
become impossible. This is well-known in GR when we consider second-order perturbations, where all the sectors of the first-order
perturbations become the sources of the second-order ones \cite{MW09}. Taking these into account, let us consider the non-perturbative solutions 
of the type,
\bqn
\lb{6.1}
N &=& a(\eta),\;\; N_{i} = a^{2}(\eta)n_{i}(t, \bf{x}),\nb\\
 g_{ij} &=& a^{2}(\eta) e^{-2\psi(t, \bf{x})}\delta_{ij}.
\eqn
After simple but tedious calculations [cf. Appendix A], we find the following exact solutions of the corresponding HL equations with a non-zero
 cosmological constant $\Lambda$,
\bqn
\lb{6.3}
a(\eta) &=& - \left(\frac{3(2-3\xi)}{2\Lambda}\right)^{1/2}\;   \frac{1}{ \eta},\nb\\
\psi &=& \psi_{0},\;\;\; n_{i} = n_0\left(- y, x, 0\right),
\eqn
where $\psi_{0}$ and $n_0$ are two integration constants. Without loss of generality, one can  set $\psi_{0} = 0$ by rescaling of the coordinates 
$x^i$ (and redefinition of  the constant $n_{0}$). 
The constant $n_{0}$, on the other hand,  represents the rotation of the spacetime, and cannot be gauged away, 
as can be seen from the following analysis. If one considers the rotation as perturbations, one can see that it corresponds to  the sum of
 infinitely high order perturbation terms, some of which will become singular in the limit $\xi \rightarrow 0$, as it is expected
 from the analysis given in the last section. But, the analytical solutions themselves
 indeed have a finite and smoothy  limit,  $\xi \rightarrow 0$, as one can see  from Eq.(\ref{6.3}). 
In particular, when $\xi = 0$, the above solutions reduce to a rotating de Sitter spacetime. In fact, introducing 
the cylindrical coordinates $r$ and $\theta$ via the relations $x = r \cos\left(\theta\right)$ and 
$y = r \sin\left(\theta\right)$, we find that the metric can be written in the form,
\bqn
\lb{6.4}
\left. ds^{2}\right|_{\xi = 0} &=& \frac{1}{(-H\eta)^{2}}\Big\{ - d\eta^{2} + dr^{2} + dz^{2} \nb\\
& & +  \left(d\theta + n_{0} d\eta\right)^{2}
\Big\},
\eqn
with $H = \sqrt{3/\Lambda}$, and $n_0$ represents the angular velocity of the rotation.

 \section{Conclusions}
\renewcommand{\theequation}{5.\arabic{equation}} \setcounter{equation}{0}

In this paper, we have considered two different issues raised recently in the studies of the HL theory, the stability of background spacetime and strong
coupling, by paying main attention on the SVW setup \cite{SVW}, which represents the most general HL theory with projectability condition.
Although the Minkowski spacetime is not stable in such a setup, the de Sitter spacetime is, due to    two different kinds 
of effects: one is from the high-order derivatives of the spacetime curvature, and the other is from the exponential expansion 
of the de Sitter space. By combining these  effects  properly, one can make  the instability found in the Minkowski background 
never raise even for small-scale modes. The condition is simply that of Eq.(\ref{4.11j}), from which we can see that if the IR limit  is  
sufficiently closed to  the  relativistic  fixed point ($\xi = 0$ or $c_{\psi} = 0$), it can be satisfied. In particular, 
at the fixed point, all the modes become stabilized. 

To stabilize the massless spin-0 graviton, another way is to invoke a Higgs-like mechanism
to give it  a mass term, $m^2_{\psi}\psi^{2}$,  in the effective action \cite{Horava2}.
Massive gravity in 4-dimensional spacetimes have been intensively
studied recently, see, for example, \cite{RT} and references therein).  Then, it can be shown that the equation for the metric perturbation
 $\psi_k$ in
 the Minkowski background 
reads, $\ddot{\psi}_{k} + \omega_{k}^{2} \psi_{k} = 0$, but now with 
\bq
\lb{5.1}
\omega^{2}_{k} = \left|c_{\psi}\right|^{2}\left(m^{2}_{\psi} -  k^{2}  + \frac{k^{4}}{M_{A}^{2}}
 + \frac{k^{6}}{M_{B}^{4}}\right).
\eq
Clearly, if $m_{\psi}$ is large enough, $\omega_{k}^{2}$ is always non-negative for any given $k$. It is not difficult to show that  
such a condition is 
\bq
\lb{5.2}
{m}_{\psi} \ge  \frac{ \; {M_{B} \sqrt{4 + r^{2}}} }{\left(\sqrt{3+ r^{4}} + r^{2}\right)^{1/2}\left(r^{4} + r^{2}\sqrt{3+ r^{4}} + 6\right)^{1/2}},
\eq
from which we find that 
$m_{\psi}(r \sim 1) \simeq M_{B}$.

The strong coupling problem has been also investigated, and found that it 
cannot be solved by the Blas-Pujolas-Sibiryakov mechanism, initially designed for the case without
projectability condition.  Strong coupling itself  is not a problem, but an indication that the linear perturbations are broken when energies involved
in processes of interest are higher than the strong coupling energy. Then,  nonlinear effects are needed to be taken into account. If the relativistic
 limit (or very closed to it) can be obtained in the IR, after the non-linear effects are taken into account, the theory is still viable. Two typical examples
 are the massive gravity  \cite{Vain} and the DGP brane model \cite{DDGV}, although the physics behind of them is different
  [See Footnote 1 given in the Introduction]. In this paper, we have constructed a
class of non-perturbative cosmological solutions in the SVW setup, and shown explicitly that it reduces smoothly to the rotating de Sitter 
spacetime.  This can be considered as a cosmological generalization of the spherical case studied recently by Mukohyama \cite{Mukc}.

~\\{\bf Acknowledgements:} We would like to thank R. Brandenberger, R.-G. Cai, B. Hu, 
K. Koyama,  R. Maartens,  S. Mukohyama,  A. Papazoglou,  O. Pujolas,  T.P. Sotiriou and 
B.F.L. Ward  for valuable discussions and comments.  Our special thanks go to A. Papazoglou 
and T.P. Sotiriou for their critical reading and suggestions. Our special thanks also go to 
O. Pujolas for his suggestive comments.  The work of AW was supported in part by DOE 
Grant, DE-FG02-10ER41692, and the one of QW by NSFC Grant, 11075141. 

\section{Appendix A: Non-linear Cosmological Perturbations}
\renewcommand{\theequation}{A.\arabic{equation}} \setcounter{equation}{0}

In this Appendix, we present some basic expressions that are useful for the studies of
 the non-linear cosmological perturbations, given by
\bqn
\lb{A.1}
N(\eta) &=& a(\eta), \;\;\; N_{i} = a^{2}(\eta)n_{i}(\eta, {\bf{x}}),\nb\\
g_{ij} &=& a^{2}(\eta)e^{-2\psi(\eta, {\bf{x}})} \delta_{ij}.
\eqn
For the sake of simplicity, we set $c = 1$. Then, we find that
\bqn
\lb{A.2}
K_{ij} &=& a\Big(e^{-2\psi}\big(\psi' - {\cal{H}}\big)\delta_{ij} + n_{(i,j)} - n_{k}\psi^{,k}\delta_{ij}\nb\\
& & + n_i\psi_{,j} + n_j\psi_{,i} \Big),\nb\\
K &=&  \frac{e^{2\psi}}{a}\left(3e^{-2\psi}\big(\psi' - {\cal{H}}\big)  -   n_{k}\psi^{,k} + \partial^{k}n_{k}\right),\nb\\
R_{ij} &=&  \psi_{,ij} + \psi_{,i}\psi_{,j} + \Big(\partial^{2}\psi - \big(\partial\psi\big)^{2}\Big)\delta_{ij},\nb\\
R &=& \frac{2e^{2\psi}}{a^{2}}\left(2\partial^{2}\psi - \big(\partial\psi\big)^{2}\right),
\eqn
where $n_{(i,j)} \equiv (n_{i,j} + n_{j,i})/2$. Then, we obtain that
\bqn
\lb{A.3a}
 {\cal{L}}_{K} &=& K_{ij}K^{ij}- \lambda K^{2}\nb\\
&=& \frac{e^{4\psi}}{a^{2}}\Big\{3\big(1-3\lambda)e^{-4\psi} \big(\psi' - {\cal{H}}\big)^{2}\nb\\
& &  
      + \big(1-\lambda\big)n_{k}\psi^{,k}\Big(n_{k}\psi^{,k} - 2\partial^{k}n_{k}\Big) \nb\\
&  & - 2 \big(1-3\lambda)e^{-2\psi} \big(\psi' - {\cal{H}}\big) \Big(n_{k}\psi^{,k} - \partial^{k}n_{k}\Big)
       \nb\\
& & + n_{(k,l)}\big(n^{(k,l)}   + 2 n^{k}\psi^{,l} + 2n^{l}\psi^{,k}\big)  \nb\\
& &   + 2 n_{k}n^{k}\psi_{,l}\psi^{,l} - \lambda\left(\partial^{k}n_{k}\right)^{2}\Big\},\\
\lb{A.3b}
R_{ij}R^{ij} &=& \frac{e^{4\psi}}{a^{4}}\Big\{5\big(\partial^{2}\psi\big)^{2} + 6 \big(\partial^{2}\psi\big)\big(\partial\psi\big)^{2}
+ 2\big(\partial\psi\big)^{4}\nb\\
& & + \psi_{kl}\big(\psi^{,kl} + 2\psi^{,k}\psi^{,l}\big)\Big\},\\
\lb{A.3c}
R^{i}_{j}R^{j}_{k}R^{k}_{i} &=&  \frac{e^{4\psi}}{a^{4}}\Big\{\psi^{,i}_{\; ,j}\psi^{,j}_{\; ,k}\psi^{,k}_{\; ,i}
+ 3 \psi^{,ij}\psi_{,jk}\psi^{,k}\psi_{,i} \nb\\
& & + 3\psi_{,ij}\psi^{,ij}\Big(\partial^{2}\psi - \big(\partial\psi\big)^{2}\Big)\nb\\
& & + 3\psi_{,ij}\psi^{,i}\psi^{j}\Big(2\partial^{2}\psi - \big(\partial\psi\big)^{2}\Big)\nb\\
& & + 6\big(\partial^{2}\psi\big)^{2} \Big(\partial^{2}\psi - 2\big(\partial\psi\big)^{2}\Big)\nb\\
& & + \big(\partial\psi\big)^{4}\Big(9\partial^{2}\psi - 2\big(\partial\psi\big)^{2}\Big),
\eqn
and
\bqn
\lb{A.4}
& & \big(\nabla_{i}R_{jk}\big)\big(\nabla^{i}R^{jk}\big) =  \frac{e^{6\psi}}{a^{6}}\Big\{F_{ijk}F^{ijk} + 3 G_{k}G^{k}\nb\\
& & ~~~~~~ + 4 F^{i}\Big(F^{k}_{\;\;\;ki} + 2F_{i}\Big) + 2G^{i}\Big(F_{ik}^{\;\;\;k} +2 F_{i}\Big), \nb\\
\eqn  
where  
\bqn
\lb{A.4a}
F_{ijk} &=& \psi_{,ijk} + 2\psi_{,ij}\psi_{,k} + 2 \psi_{,ik}\psi_{,j} + 2 \psi_{,jk}\psi_{,i} \nb\\
& &  + 4\psi_{,i} \psi_{,j} \psi_{,k},\nb\\
G_{i} &=& \Big(\partial^{2}\psi - \big(\partial\psi\big)^{2}\Big)_{,i} +
2 \Big(\partial^{2}\psi - \big(\partial\psi\big)^{2}\Big)\psi_{,i}, \nb\\
F_{i} &=& - \Big(\psi^{,k}_{\;\;,i} + \delta^{k}_{i}\big(\partial\psi\big)^{2}\Big)\psi_{,k},
\eqn
and $F^{i} \equiv \delta^{ik}F_{k}$, etc. 

With these expressions, one can write down the Hamiltonian and momentum constraints,
and the dynamical equations given in \cite{WM,WWM}, which are too complicated to provide here.


\end{document}